\documentclass[%
    aps,
    prl,
    10pt,
    nofootinbib,
    noeprint,
    amsmath,
    amssymb,
    preprintnumbers,
    superscriptaddress,
    twocolumn
]{revtex4-2}

\usepackage{graphicx}
\usepackage[%
    colorlinks=true,
    allcolors=blue
]{hyperref}
\usepackage[%
    print-unity-mantissa=false,
    range-phrase=-,
    range-units=single
]{siunitx}

\DeclareSIUnit\parsec{pc}

\def\actaa{Acta\ Astronaut.}
\def\baas{Bull.\ Am.\ Astron.\ Soc.}
\def\cqg{Classical\ Quant.\ Grav.}
\def\ijmpe{Int.\ J.\ Mod.\ Phys.\ E}
\def\na{Nat.\ Astron.}
\def\nc{Nat.\ Commun.}
\def\np{Nat.\ Phys.}
\def\mnras{Mon.\ Not.\ R.\ Astron.\ Soc.}
\def\physrep{Phys.\ Rept.}
\def\prr{Phys.\ Rev.\ Res.}
\def\prx{Phys.\ Rev.\ X}

\begin{document}

\title{Interface modes in inspiralling neutron stars: \\A gravitational-wave probe of first-order phase transitions}

\author{A. R. Counsell}
\email{a.r.counsell@soton.ac.uk}
\affiliation{Mathematical Sciences and STAG Research Centre, University of Southampton, Southampton SO17 1BJ, United Kingdom}

\author{F. Gittins}
\email{f.w.r.gittins@uu.nl}
\affiliation{Institute for Gravitational and Subatomic Physics (GRASP), Utrecht University, Princetonplein 1, 3584 CC Utrecht, Netherlands}
\affiliation{Nikhef, Science Park 105, 1098 XG Amsterdam, Netherlands}

\author{N. Andersson}
\affiliation{Mathematical Sciences and STAG Research Centre, University of Southampton, Southampton SO17 1BJ, United Kingdom}

\author{I. Tews}
\affiliation{Theoretical Division, Los Alamos National Laboratory, Los Alamos, New Mexico 87545, USA}

\preprint{LA-UR-25-22199, INT-PUB-25-008}
\date{\today}

\begin{abstract}
At the extreme densities in neutron stars, a phase transition to deconfined quark matter is anticipated.
Yet masses, radii and tidal deformabilities offer only indirect measures of a first-order phase transition, requiring many detections to resolve or being ineffective observables if the discontinuity exists at lower densities.
We report on a \textit{smoking-gun} gravitational-wave signature of a first-order transition: the resonant tidal excitation of an interface mode.
Using relativistic perturbation theory with an equation-of-state family informed by chiral effective field theory, we show that such a resonance may be detectable with next-generation interferometers and possibly already with LIGO A+ for sufficiently loud events.
\end{abstract}

\maketitle

\textit{Context}.---%
Neutron stars are Nature's cosmic laboratory for dense matter, where densities far exceed those that can be reached in terrestrial experiments. 
Understanding the properties of matter at such extreme conditions remains an open problem in nuclear physics. 
At low densities, nuclear interactions can be systematically described using chiral effective field theory (EFT)~\cite{2009RvMP...81.1773E,2011PhR...503....1M}, which provides an order-by-order expansion based on symmetries of quantum chromodynamics (QCD).
Chiral EFT is reliable up to approximately $\numrange{1}{2}$ times the nuclear-saturation density $n_\text{sat} \approx \qty{0.16}{\per\femto\metre\cubed}$ \cite{2018ApJ...860..149T,2020PhRvC.102e4315D}.
Neutron-star cores, however, reach central densities of a few up to $\sim 8 n_\text{sat}$, where theoretical uncertainties remain significant.
A particularly intriguing possibility is that, at sufficiently high densities, hadronic matter undergoes a phase transition to deconfined quark matter.
This transition is a robust prediction of QCD, but its exact nature---whether it is a smooth crossover or a first-order transition with a sharp interface---remains unresolved.
While collider experiments at RHIC may have provided tentative evidence for critical behaviour (see Ref.~\cite{2024IJMPE..3330008D} and their Fig.~1), the results are far from conclusive. 
Astrophysical neutron-star observations present an opportunity to probe this phase transition, should they harbour it.

To date, the properties of dense nuclear matter have primarily been explored by measuring the mass $M$, radius $R$ and tidal deformability $\Lambda$ of neutron stars.
Recent advances in observations have placed increasingly stringent constraints on these parameters. 
In particular, the X-ray timing mission NICER has provided mass-radius measurements for a few pulsars \cite{2019ApJ...887L..21R,2019ApJ...887L..22R,2019ApJ...887L..24M,2021ApJ...918L..27R,2021ApJ...918L..28M,2021ApJ...918L..29R}. 
Meanwhile, gravitational-wave data, particularly from the landmark GW170817 event, have placed upper limits on the neutron-star tidal deformability \cite{2017PhRvL.119p1101A,2018PhRvL.121i1102D,2018PhRvL.121p1101A,2019PhRvX...9a1001A}.
Combined, these observations suggest error bars of $O(\qty{1}{\kilo\metre})$ for the neutron-star radius.
The next generation of gravitational-wave observatories---Cosmic Explorer \cite{2019BAAS...51g..35R} and the Einstein Telescope \cite{2010CQGra..27s4002P}---is expected to enhance our ability to measure these aspects, providing much tighter constraints on the high-density physics \cite{2022PhRvD.105h4021C}.

\begin{figure*}
    \includegraphics[width=\columnwidth]{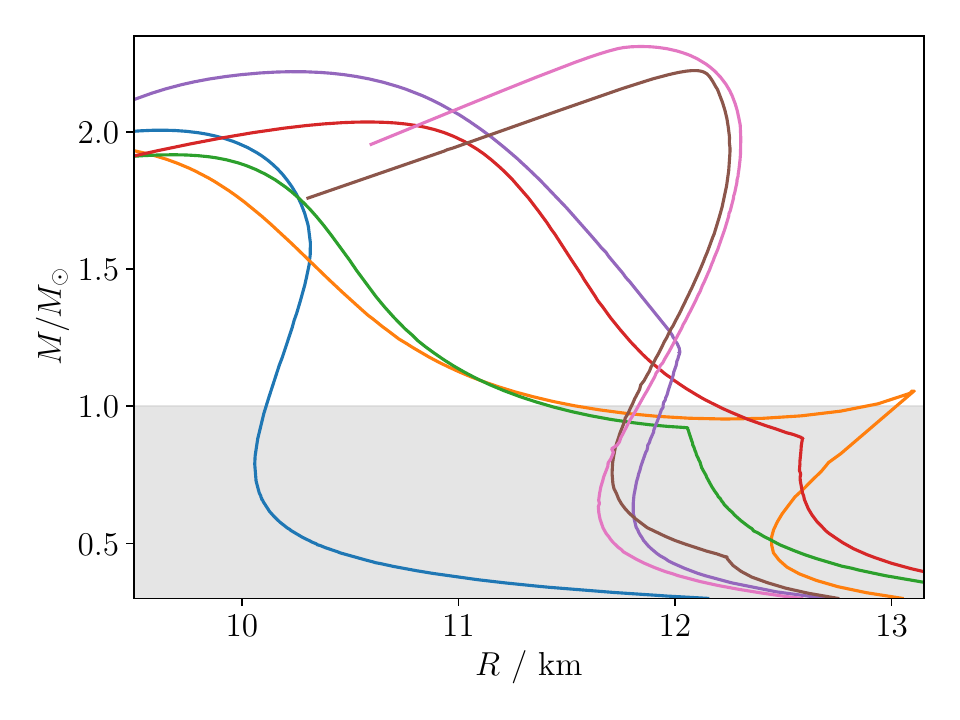}%
    \includegraphics[width=\columnwidth]{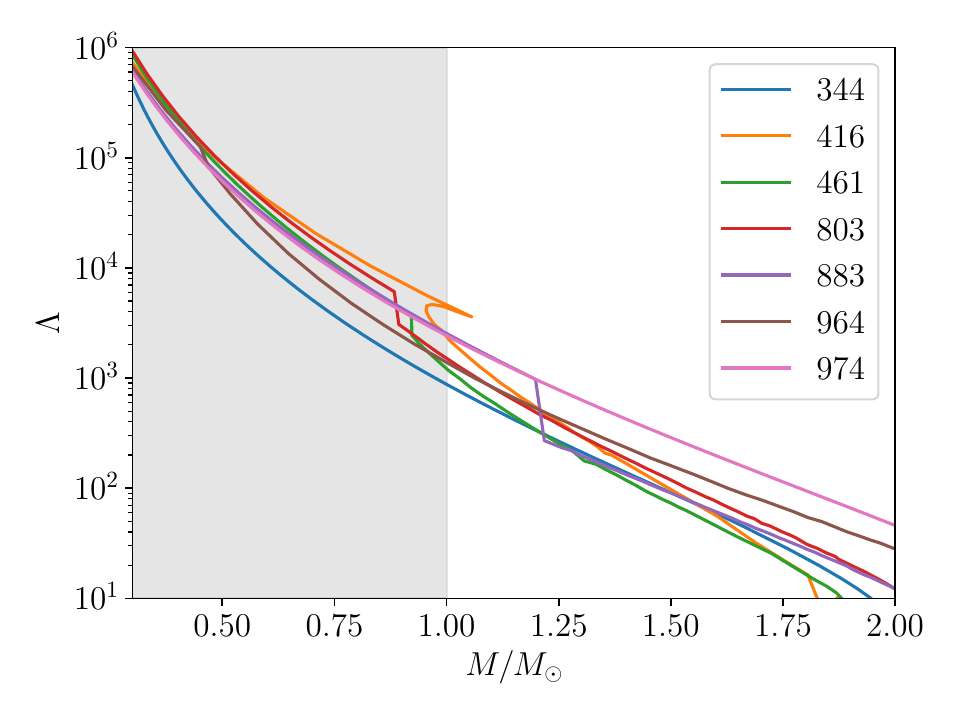}
    \caption{\label{fig:MRCurves}%
    The mass-radius (left panel) and tidal deformability curves (right panel) for a selection of EOS models from Ref.~\cite{2020NatAs...4..625C}, indexed by their ordering in radius at $M = 1.4 M_\odot$.
    Five of these matter models possess first-order phase transitions---manifesting as kinks in the $M - R$ plot---while two do not.
    For the ensemble of 2000 models we consider here, the majority of phase transitions occur at sufficiently low densities such that their impact is only visible below $M = M_\odot$, as indicated by the shaded regions.}
\end{figure*}

When it comes to the issue of unveiling the presence of a phase transition, it is important to note that $M$, $R$ and $\Lambda$ reflect global, averaged characteristics of the star.
In particular, the tidal deformability $\Lambda$ represents the star's susceptibility to an external gravitational field in the portion of the inspiral where the compact binary is well separated and the tidal field may be approximated as static \cite{2008PhRvD..77b1502F,2008ApJ...677.1216H}.
While these parameters do contain information about possible phase transitions, the relevant features may be masked or degenerate with other aspects of the equation of state (EOS; see, e.g., Refs.~\cite{2005ApJ...629..969A,2019PhRvD..99h3014H,2023PhRvL.130t1403R,2023PhRvD.108b3010R}).
We illustrate the issue in Fig.~\ref{fig:MRCurves} for a subset of the EOS models we consider later.
These models are consistent with chiral EFT up to nuclear saturation and then extrapolated to higher densities in such a way that each model supports the existence of $1.9 M_\odot$ neutron stars, in agreement with heavy-pulsar observations. No other constraints were imposed (see Refs.~\cite{2018PhRvC..98d5804T,2020NatAs...4..625C} for more details).
Some of these models include a phase transition.
Figure~\ref{fig:MRCurves} displays $M$, $R$ and $\Lambda$, highlighting the difficulty of unambiguously identifying a phase transition from these bulk observables alone.

If the transition is of first order, the $M - R$ curve will exhibit a kink or jump, associated with an effective softening of the matter.
In the $\Lambda - M$ curve, the transition manifests as a discontinuity over a small range in mass, leading to a sharp decrease in $\Lambda$; the higher the pressure at the transition, the less $\Lambda$ decreases \cite{2019PhRvD..99h3014H}.
Above the transition pressure, there are no distinguishable features and the curves behave as one would expect for a purely hadronic EOS (compare, e.g., the two models labelled 964 and 974 in the region $\numrange{1}{2} M_\odot$ in Fig.~\ref{fig:MRCurves}).
This is a particular manifestation of the so-called \textit{masquerade problem} \cite{2005ApJ...629..969A}.

For several of the models we consider, the phase transitions occur at masses $\lesssim M_\odot$.
Assuming the standard supernova mechanism of generating neutron stars, recent work suggests a minimum formation mass of $\sim 1.17 M_\odot$ \cite{2018MNRAS.481.3305S}.
Therefore, the phase-transition region is unlikely to be probed by neutron-star measurements.
In order to identify the phase transition, one would need it to occur at sufficiently high mass and resolve the region around that transition mass to high enough precision.
Assuming an exceptional GW170817-like event observed by third-generation detectors, one might be able to constrain the error in the neutron-star radius $\Delta R$ to $\sim \qtyrange{50}{200}{\metre}$ and tidal deformability to a level of $\Delta{\Lambda} \sim 20$ for a range of masses~\cite{2022PhRvD.105h4021C}.
This ideal scenario, ignoring possible systematic errors~\cite{{boudon2025realisticprojectionconstrainingneutron}}, would be enough to constrain the matter models in Fig.~\ref{fig:MRCurves}, provided the range of observations included the transition mass.
Reaching the required level of precision for weaker events will be challenging.

There have, nevertheless, been valuable attempts to use measurements of $M$, $R$ and $\Lambda$ to search for the presence of phase transitions.
Recent studies~\cite{2020NatPh..16..907A,2023NatCo..14.8451A} argue that the most massive neutron stars contain quark-matter cores.
Additionally, it has been shown that gravitational waves may provide evidence for strong phase transitions \cite{2019PhRvD..99h3014H,2020PhRvD.101d4019C,2020PhRvR...2c3514P}.
In this Letter, we concern ourselves with a dynamical effect in binary inspirals that may provide additional information about the transition.
The main goal is to highlight an observational strategy that could provide direct evidence for a sharp, first-order phase boundary.

\textit{The dynamical tide}.---%
As a compact binary evolves during the inspiral, the tidal frequency eventually increases to a level that is comparable to the hydrodynamical timescale of the neutron star. 
At this point, the static assumption breaks down and the tide becomes dynamical, implicating the neutron star's vibrational modes. 
As the tidal frequency rises due to gravitational-wave emission, it may momentarily match the frequency of a natural oscillation, leading to a resonance where the mode amplitude rapidly grows as energy is extracted from the orbit. 
This abrupt removal of energy advances the orbital decay and leaves an imprint on the gravitational-wave phase.

In the Newtonian limit, the contribution of an individual mode to the orbital motion is determined by two key quantities: its oscillation frequency $\omega$ and tidal overlap integral $Q_l$, where $l$ is the multipolar order of the perturbation.
The mode frequency dictates the point in the inspiral at which resonance occurs, while the tidal overlap quantifies how efficiently the tide couples to the mode.
Previous studies have paid particular attention to the tidal excitation of compositional \textit{g}-modes \cite{1994ApJ...426..688R,1994MNRAS.270..611L,1995MNRAS.275..301K,2017MNRAS.464.2622Y,2017MNRAS.470..350Y,2021MNRAS.506.2985K,2025MNRAS.536.1967C}, which are supported by chemical-composition gradients in the star.
This body of work has found that---although the \textit{g}-mode frequencies lie within the inspiral band---the (quadrupolar) overlaps of $Q_2 / (M R^2) \sim \numrange{e-5}{e-4}$ may be too weak to produce a measurable effect on the binary's gravitational-wave phase (see Refs.~\cite{2023PhRvD.108d3003H,2025MNRAS.536.1967C} for recent estimates).

\textit{Lessons from an incompressible stellar model}.---%
Here, we explore a related family of buoyancy modes that have significantly more favourable prospects for detection and provide a more direct probe of high-density phase transitions: the interfacial \textit{i}-modes (also known as \textit{discontinuity} \textit{g}-modes) \cite{1979AcA....29..135G,1987MNRAS.227..265F,1990MNRAS.245..508M,1993ApJ...417..273S,2003MNRAS.338..389M}.
Suppose the neutron star has a sharp density discontinuity in its interior and a fluid element traverses this interface.
The fluid parcel, which maintains pressure equilibrium with its surroundings, has the properties of its original position.
As it is displaced across the interface, the element will suddenly be in an environment of differing composition and density.
Assuming that the reaction timescales that equilibrate the fluid element to its new environment are \textit{slower} than the timescale of the perturbation, it will be subject to a buoyancy force that restores the parcel to its origin on the other side of the interface
(analogous to compositional \textit{g}-modes \cite{1992ApJ...395..240R}). This is expected to be the case if the relevant phase conversion (the speed of the phase boundary) is limited by the weak interaction (as in the model explored in Ref.~\cite{2015PhRvC..91e5804A}). If, on the other hand, the strong-interaction timescale applies then the star will not support the kind of interface modes we consider here.

A simple, analytical calculation of an incompressible fluid sphere with a sharp interface (detailed in the Supplemental Material) shows that the frequency of an \textit{i}-mode is approximated by
\begin{equation}
\begin{split}
    \omega^2 \approx (2 \pi \times \qty{686}{\hertz})^2 &\left( \frac{\epsilon}{0.1} \right) \left( \frac{M}{1.4 M_\odot} \right) \left( \frac{\qty{10}{\kilo\metre}}{R} \right)^3 \\
    &\times \frac{l (l + 1)}{2 l + 1} \left[ 1 - \left( \frac{r_\text{i}}{R} \right)^{2 l + 1} \right]\,,
\end{split}
    \label{eq:ToyModel}
\end{equation}
where $r_\text{i}$ is the radial position of the interface in the star and $\epsilon$ is the relative jump in mass density. 
We see that $\omega^2$ depends on the location of the interface and linearly on the jump, meaning that an observed \textit{i}-mode could provide a direct measurement of the discontinuity.
For the assumed scalings, the mode falls within the frequency band of a typical binary inspiral.

This computation demonstrates why the \textit{i}-modes may be promising from the observational point of view.
We find that the \textit{i}-mode eigenfunctions bear a striking resemblance to those of the fundamental \textit{f}-mode; the oscillation mode that couples the most strongly with tides \cite{2020PhRvD.101h3001A}:
the perturbation grows as $\propto r^l$ from the stellar centre to the interface $r_\text{i}$, which mimics the functional dependence of the tidal driving force.
However, above the interface $r > r_\text{i}$, the eigenfunctions alter and obtain a non-vanishing decaying contribution, giving rise to a characteristic sharp kink at the interface.
From the mode solution, we determine the tidal overlap to be (with normalisation $\mathcal{A}^2 = M R^2$)
\begin{equation}
\begin{split}
    Q_l = &\num{-1e-2} \left( \frac{\epsilon}{0.1} \right)^2 M R^l \sqrt{\frac{3}{4 \pi} \frac{l (l + 1)}{2 l + 1}} \\
    &\times \sqrt{1 - \left( \frac{r_\text{i}}{R} \right)^{2 l + 1}} \left( \frac{r_\text{i}}{R} \right)^{(2 l + 1) / 2} \left[ 1 - \left( \frac{r_\text{i}}{R} \right)^3 \right]\,.
\end{split}
    \label{eq:QToyModel}
\end{equation}
This shows that (for modest values of the discontinuity) the tidal coupling is at least an order of magnitude larger than the aforementioned \textit{g}-mode estimates, suggesting that the \textit{i}-modes can couple quite efficiently to the tidal forcing and may therefore be observable in binary inspirals.

Interfaces, such as at the crust-core boundary \cite{2012PhRvL.108a1102T,2020PhRvL.125t1102P,2023PhRvD.107h3023Z}, naturally exist in neutron stars.
Previous work has focused on tidally induced crust fractures and possible electromagnetic precursor signals for binary neutron-star mergers.
Here, our focus is on the prospects of using \textit{i}-modes to reveal the potential phase transition to deconfined quarks in the neutron-star core \cite{2021PhRvD.103f3015L,2024ApJ...964...31M}.
If such a transition is first order, it will generate an \textit{i}-mode signature in the gravitational-wave signal from an inspiralling neutron-star binary.
To answer how significant this signature is, we require a relativistic calculation with realistic microphysics.

\textit{A family of equations of state}.---%
We assume a barotropic EOS $p = p(\varepsilon)$, relating the pressure $p$ to the energy density $\varepsilon$ of the fluid.
This has two advantages.
First, the star will not support composition \textit{g}-modes, which simplifies the mode spectrum we have to consider.
Second, treating the matter as a perfect fluid, we can immediately calculate the speed of sound $c_\text{s}$ as measured by a co-moving observer using
\begin{equation}
    \left( \frac{c_\text{s}}{c} \right)^2 = \frac{d p}{d \varepsilon}\,,
    \label{eq:SpeedOfSound}
\end{equation}
where $c$ is the speed of light. 
We adopt $c_\text{s}^2$ as the fundamental thermodynamical variable and integrate Eq.~\eqref{eq:SpeedOfSound} to obtain $p = p(\varepsilon)$.
Invoking causality and thermodynamical stability arguments leads to the trivial constraint $0 \leq (c_\text{s} / c)^2 < 1$.
Using results from low densities, where the theoretical understanding is robust, one can extend the EOS to higher densities by sampling $c_\text{s}^2$ and ensuring that it obeys the constraints \cite{2018ApJ...860..149T,2018PhRvC..98d5804T}.
This sampling generates an EOS family.

One may generate additional models by allowing for first-order phase transitions of arbitrary location and width. 
This approach does not allow us to extract information on the composition of dense matter or the type of a phase transition but it does provide a useful way to test the effect such a transition has on neutron-star properties and observables.
Moreover, the collection of models we use---the same as in Ref.~\cite{2020NatAs...4..625C}---is agnostic in the sense that some of the models have phase transitions, while others do not.
Indeed, those that do not will have no associated interface mode.

\textit{Interface modes}.---%
To make progress beyond the incompressible model, we need to use general relativity.
For simplicity, we ignore temperature, rotation and continue to assume a perfect-fluid star.
As the corresponding linear perturbation problem has been explored in great detail in the literature and the various steps required in its formulation are well known, we only outline the strategy here.
Our calculation follows the steps laid out in Ref.~\cite{2025MNRAS.536.1967C}; we adopt the relativistic Cowling approximation (ignore perturbations of the spacetime metric and hence do not include gravitational-wave emission associated with the modes). 
Unlike previous calculations of interface modes, such as Ref.~\cite{2024ApJ...964...31M}, we do not include any jump conditions or perform matching at the phase transition when solving the oscillation mode equations. 
Instead, we take an agnostic approach and solve the equations in an identical manner regardless of whether or not a phase transition is present, allowing the numerics to identify any discontinuities.

To quantify the impact of a resonance on the orbital motion, we require the overlap integral $Q_l$.
In the Cowling approximation used here, this is given by \cite{2025MNRAS.536.1967C}
\begin{equation}
\begin{split}
    Q_l &= \frac{1}{c^2} \int (\varepsilon + p) \xi^{i*} \nabla_{i}(r^l Y_{lm}) \sqrt{-g} \, d^3x \\
    &= \frac{1}{c^2} l \int_0^R e^{(\nu + \lambda)/2} (\varepsilon + p) r^l \left[ W_l + (l + 1) V_l \right] \, dr\,,
\end{split}
\end{equation}
where $W_l$ and $V_l$ represent the radial and angular components of the Lagrangian fluid displacement $\xi^i$, respectively, $\nu$ and $\lambda$ are functions in the metric $g_{a b}$ with determinant $g = \text{det}(g_{a b})$ and $Y_l^m$ is a spherical harmonic of degree $l$ and order $m$.
To normalise our results, we introduce
\begin{equation}
\begin{split}
    \mathcal{A}^2 &= \frac{1}{c^2} \int e^{-\nu} (\varepsilon + p) \xi^{i*} \xi_i \sqrt{-g} \, d^3x \\
    &= \frac{1}{c^2} \int_0^R e^{(\lambda - \nu)/2} (\varepsilon + p) \left[ e^\lambda W_l^2 + l (l + 1) V_l^2 \right] \, dr\,.
\end{split}
\end{equation}
We provide additional details about this calculation and present the eigenfunctions $W_l$ and $V_l$ for a typical mode in the Supplemental Material.

\textit{Resonance detectability}.---%
With the mode calculation in hand, we want to quantify the extent to which the associated tidal resonance in a binary inspiral signal may be detectable.
However, as the problem of tidal resonances is not yet completely formulated in general relativity, we resort to an approximation based on the Newtonian tidal interaction.
The resonance occurs between the mode frequency $\omega$ and the orbital frequency $\Omega$ when (for modes with $|m| = 2$) $\omega \approx 2\Omega = 2\pi f$, where $f$ is the gravitational-wave frequency.
The shift in orbital phase $\Delta \Phi$ due to the energy transfer to a mode during the inspiral of a binary with primary mass $M$ and component mass $M'$ can be estimated as \cite{1994MNRAS.270..611L} (see also Ref.~\cite{tidyn_paper})
\begin{equation}
\begin{split}
    \frac{\Delta \Phi}{2 \pi} \approx -& \frac{5 \pi}{4096} \left( \frac{c^2 R}{G M} \right)^5 \frac{2}{q (1 + q)} \\
    &\times \frac{G M / R^3}{\omega^2} \left( \frac{Q_l}{M R^l} \right)^2 \frac{M R^2}{\mathcal{A}^2} \,,
\end{split}
    \label{eq:ModeShift}
\end{equation}
where $q = M' / M$ is the mass ratio.
We assume that a similar expression applies in the relativistic case---although this admittedly remains to be justified by a detailed derivation---with the scaling based on the fully relativistic results for the star's mass and radius (see Refs.~\cite{2023PhRvD.108d3003H,2025MNRAS.536.1967C} for similar estimates).

\begin{figure}
    \includegraphics[width=\columnwidth]{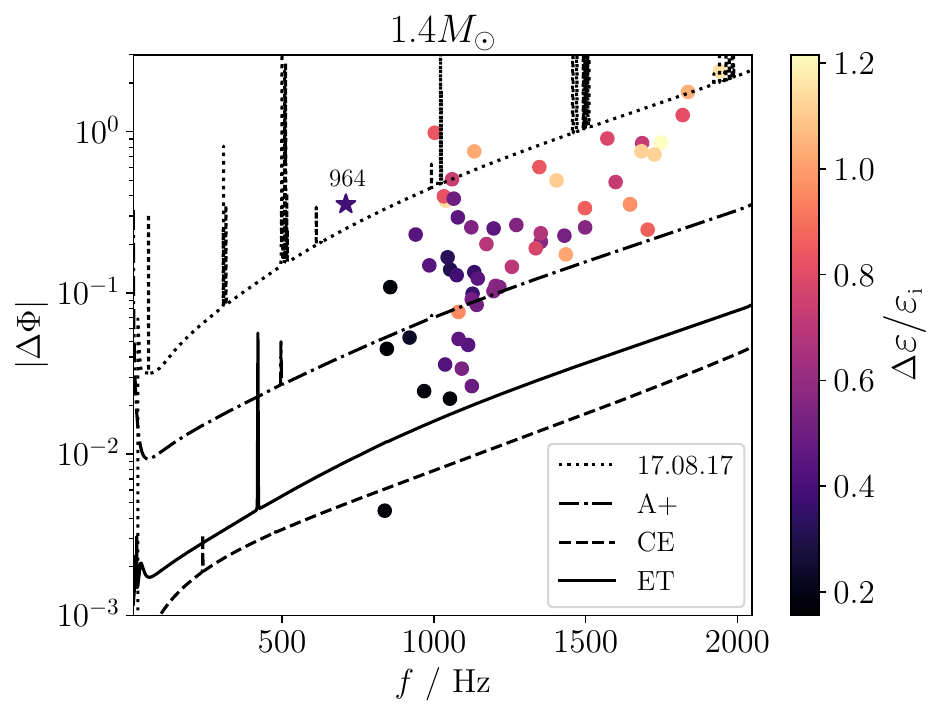}%
    \caption{\label{fig:DetectabilityTrendsEps}%
    Estimated shift in orbital phase $|\Delta \Phi|$ against gravitational-wave frequency $f$ for an equal-mass $M = 1.4 M_\odot$ binary.
    Each marker corresponds to an $l = 2$ \textit{i}-mode resonance computed from a different EOS in the ensemble, coloured by the relative jump in energy density $\Delta \varepsilon / \varepsilon_\text{i}$.
    We indicate with a star the interface mode associated with model 964 in Fig.~\ref{fig:MRCurves}.
    Also inlaid are the sensitivity curves for LIGO Livingston during the GW170817 event (dotted line), LIGO A+ (dot-dashed line), Cosmic Explorer (CE, dashed line) and the Einstein Telescope (ET, solid line), assuming the binary is at a luminosity distance of \qty{40}{\mega\parsec} from the instrument.}
\end{figure}

It is, however, important to add caveats to these assumptions. In particular,  it is clear that some of the interface modes shown in Fig.~\ref{fig:DetectabilityTrendsEps} have frequencies such that they would only become resonant very late in the binary inspiral.
Based on the estimated merger frequency (from Ref.~\cite{2013PhRvD..88d4042R}), we find that all the resonances indicated in Fig.~\ref{fig:DetectabilityTrendsEps} occur within the last 5 orbits before merger.
In this regime the Newtonian inspiral estimate we are relying on is---at best---indicative as it ignores the impact of the curved spacetime and nonlinear aspects of the tide (see Ref.~\cite{2023MNRAS.519.4325Y} for a relevant discussion).
However, given the lack of fully relativistic models for the dynamical tide (including actual resonances) and the challenges associated with developing such models \cite{2024PhRvD.109f4004P}, we have to resort to this approximate analysis. 
Our expectation is that our proof-of-principle demonstration will stimulate the development of truly quantitative tidal models along with parallel work on the problems of detection and parameter inference.

To assess detectability, we must quantify the minimum observable shift in orbital phase $\Delta \Phi(f)$ as a function of $f$ for a given interferometer.
This can be estimated from \cite{2023CQGra..40m5002R}
\begin{equation}
    |\Delta \Phi(f)| = \frac{\sqrt{S_n(f)}}{2 A(f) \sqrt{f}} \,,
\end{equation}
where $S_n(f)$ is the noise power spectral density of the chosen detector and $A(f)$ is the gravitational-wave amplitude of the waveform.
For simplicity, we consider only the \verb+IMRPhenomPv2_NRTidal+ waveform model, which includes the static tidal deformabilities, but not the dynamical contribution from the \textit{f}-mode \cite{2019PhRvD.100d4003D}.
As our aim is to estimate the detectability of the interface-mode resonances---at a level of accuracy at which different waveform models agree---with measured sensitivities of current and expected sensitivities of future gravitational-wave detectors.

Using our collection of EOS models, the $l = 2$ interface modes were calculated for a set of neutron-star masses. 
We then used the obtained values for $\Delta \Phi$ from Eq.~\eqref{eq:ModeShift} for an equal-mass system at a luminosity distance of \qty{40}{\mega\parsec} (the inferred distance to GW170817) to estimate the detectability of each resonance.
Results for $M = 1.4 M_\odot$ are summarised by Fig.~\ref{fig:DetectabilityTrendsEps}.
The results show that the frequency of the mode typically lies between that of the \textit{f}-mode, $O(\qty{2}{\kilo\hertz})$, and the frequency range where one would expect to find the composition \textit{g}-modes of non-barotropic models, $O(\qty{100}{\hertz})$ (see, e.g., Ref.~\cite{2025MNRAS.536.1967C}).
This makes the interface-mode resonance distinct from the rest of the oscillation spectrum.
The results for $\Delta \Phi$ are compared to the sensitivity curve of LIGO Livingston during the observation of GW170817 and the anticipated sensitivities of LIGO A+, Cosmic Explorer and the Einstein Telescope.

The message here is clear: The majority of the identified interface-mode resonances may be detectable already by an instrument at the LIGO A+ level and (nearly) all of them would be within reach of next-generation detectors.
In accordance with the analytical calculation that led to Eqs.~\eqref{eq:ToyModel} and \eqref{eq:QToyModel}, we see that generally the larger the relative energy-density jump, the greater the mode frequency and the phase shift.
This agrees with the (obvious) expectation that detection prospects are more favourable for stronger phase transitions.
For higher masses, despite the phase-shift decrease with $M$ in Eq.~\eqref{eq:ModeShift}, our results show that approximately half of the models would still be detectable by next-generation instruments for equal-mass $M = 1.8 M_\odot$ systems (see the Supplemental Material).

It is worth emphasising that the interface modes may provide access to a lower density portion of the EOS. 
For example, in Fig.~\ref{fig:DetectabilityTrendsEps} we highlight the \textit{i}-mode resonance of model 964. 
As is evident from Fig.~\ref{fig:MRCurves}, the  phase transition in this EOS would be \textit{inaccessible} with measurements of $M$, $R$ and $\Lambda$.

While a single resonance signature would identify the presence of the phase transition, the next challenge would be to constrain the onset density and size of the energy-density jump. 
As the frequency and phase shift depend on both parameters, multiple detections of the \textit{i}-modes for different masses may be able to constrain both.

\textit{Conclusions}.---%
A robust prediction of the theory of QCD is that high-density matter undergoes a phase transition from ordinary hadronic matter to deconfined quarks.
So far, this transition has been explored with astrophysical measurements of the bulk properties of neutron stars; their masses $M$, radii $R$ and tidal deformabilities $\Lambda$.
However, obtaining evidence for a phase transition from these quantities will likely require third-generation sensitivities and data from a number of inspiral events \cite{2020PhRvR...2c3514P,2022PhRvD.105h4021C}.

In this Letter, we have explored a separate, dynamical signature of first-order phase transitions: the resonant tidal excitation of interfacial \textit{i}-modes.
The detection of an interface mode in a coalescing neutron-star binary would be a \textit{smoking-gun} signature of a first-order phase transition, distinguishable in a single gravitational-wave event.
We demonstrated the promise of the idea with results from a general-relativistic perturbation calculation using a family of nuclear-matter equations of state generated from chiral EFT.
Our estimates show that an \textit{i}-mode resonance may be observable with Cosmic Explorer and the Einstein Telescope, and possibly already with LIGO A+ for sufficiently loud gravitational-wave events.

\begin{acknowledgments}
\textit{Acknowledgements}.---%
F.G. acknowledges funding from the European Union’s Horizon Europe research and innovation programme under the Marie Sk{\l}odowska-Curie Grant Agreement No.~101151301.
N.A. gratefully acknowledges support from the STFC via Grant No.~ST/Y00082X/1.
I.T. was supported by the U.S. Department of Energy, Office of Science, Office of Nuclear Physics, under Contract No.~DE-AC52-06NA25396, by the U.S. Department of Energy, Office of Science, Office of Advanced Scientific Computing Research, Scientific Discovery through Advanced Computing (SciDAC) NUCLEI program, and by the LDRD program of LANL under Project No.~20230315ER.
N.A. and I.T thank the Institute for Nuclear Theory at the University of Washington for its kind hospitality and stimulating research environment. 
This research was supported in part by the Institute's U.S. Department of Energy Grant No.~DE-FG02-00ER41132.
\end{acknowledgments}

\textit{Data availability}.---%
The data that support the findings of this Letter are openly available \cite{code}.

\nocite{2014PhRvD..90l4023C}

\bibliography{refs}

\end{document}


\title{Supplemental Material: \\Interface modes in inspiralling neutron stars: \\A gravitational-wave probe of first-order phase transitions}

\author{A. R. Counsell}
\affiliation{Mathematical Sciences and STAG Research Centre, University of Southampton, Southampton SO17 1BJ, United Kingdom}

\author{F. Gittins}
\affiliation{Institute for Gravitational and Subatomic Physics (GRASP), Utrecht University, Princetonplein 1, 3584 CC Utrecht, Netherlands}
\affiliation{Nikhef, Science Park 105, 1098 XG Amsterdam, Netherlands}

\author{N. Andersson}
\affiliation{Mathematical Sciences and STAG Research Centre, University of Southampton, Southampton SO17 1BJ, United Kingdom}

\author{I. Tews}
\affiliation{Theoretical Division, Los Alamos National Laboratory, Los Alamos, NM 87545, United States}

\maketitle

\section{\label{app:Model}Spectrum of an incompressible star}

In order to interpret the numerical results for realistic neutron-star models---especially the scaling with the parameters associated with a phase transition---we examine the spectrum of a Newtonian, two-layer, incompressible star with a sharp density discontinuity between the layers.

We endow the star with a mass-density profile of the form
\begin{equation}
    \rho(r) = (\rho_\text{i} - \rho_\text{o}) \Theta(r_\text{i} - r) + \rho_\text{o} \Theta(R - r)\,,
\end{equation}
where $\Theta(r)$ represents the Heaviside step function, $\rho_\text{i}$ is the density of the inner sphere that extends to radius $r_\text{i}$, $\rho_\text{o}$ is the density of the outer shell and $R$ is the total radius of the star. The discontinuity occurs at $r = r_\text{i}$ and corresponds to a jump of $\rho_\text{i} - \rho_\text{o}$. We will see later how we require this density jump to be positive in order for an interface mode to manifest.

We consider perturbations about this equilibrium and treat them as incompressible, such that the Lagrangian variation of the density vanishes, $\Delta \rho = 0$. This then leads to
\begin{equation}
    \nabla_j \xi^j = 0
    \label{eq:Divergence}
\end{equation}
and
\begin{equation}
    \delta \rho = \xi^r [(\rho_\text{i} - \rho_\text{o}) \delta(r_\text{i} - r) + \rho_\text{o} \delta(R - r)]\,,
    \label{eq:deltarho}
\end{equation}
where $\delta(r)$ is the Dirac delta function, $\xi^j$ is the Lagrangian displacement vector and $\delta \rho$ is the Eulerian change of the density. Since the displacement is divergence free, Eq.~\eqref{eq:Divergence}, the fluid flow is irrotational and we can freely define a scalar potential $\psi$ such that
\begin{equation}
    \xi^j = \nabla^j \psi
\end{equation}
and immediately infer that it must satisfy Laplace's equation,
\begin{equation}
    \nabla^2 \psi = 0\,.
    \label{eq:Laplacepsi}
\end{equation}

The governing equation for the pulsations is the Euler equation. For simplicity, we will adopt the Cowling approximation and search for harmonic solutions of frequency $\omega$. Thus, we have
\begin{equation}
    - \omega^2 \xi_j = - \frac{1}{\rho} \nabla_j \delta p + \frac{\nabla_j p}{\rho^2} \delta \rho\,,
    \label{eq:Euler}
\end{equation}
where we encounter the pressure $p$ and its corresponding Eulerian perturbation $\delta p$. According to Eq.~\eqref{eq:deltarho}, $\delta \rho$ vanishes everywhere except at the two interfaces---inside the star at $r = r_\text{i}$ and at the surface $r = R$. Hence, the linearised Euler equation~\eqref{eq:Euler} reveals that $\delta p$ also satisfies Laplace's equation,
\begin{equation}
    \nabla^2 \delta p = 0 \quad \text{for} \ 0 \leq r < r_\text{i} \ \text{and} \ r_\text{i} < r < R\,,
    \label{eq:Laplacedeltap}
\end{equation}
away from the interfaces.

At this point, we know that $\psi$ and $\delta p$ obey Laplace's equation, Eqs.~\eqref{eq:Laplacepsi} and \eqref{eq:Laplacedeltap}, in the regions of the star away from the interfaces. The solutions are given simply by
\begin{align}
    \psi(r, \theta, \varphi) &= \psi_l(r) Y_l^m(\theta, \varphi)\,, \\
    \delta p(r, \theta, \varphi) &= \delta p_l(r) Y_l^m(\theta, \varphi)\,,
\end{align}
where $Y_l^m$ is a spherical harmonic of degree $l$ and order $m$, and
\begin{align}
    \psi_l(r) &= 
    \begin{cases}
        a_l r^l + b_l / r^{l + 1} \quad &\text{for} \ 0 \leq r < r_\text{i}\,, \\
        \bar{a}_l r^l + \bar{b}_l / r^{l + 1} \quad &\text{for} \ r_\text{i} < r < R\,,
    \end{cases} \\
    \delta p_l(r) &= 
    \begin{cases}
        c_l r^l + d_l / r^{l + 1} \quad &\text{for} \ 0 \leq r < r_\text{i}\,, \\
        \bar{c}_l r^l + \bar{d}_l / r^{l + 1} \quad &\text{for} \ r_\text{i} < r < R\,.
    \end{cases}
\end{align}
Since the background is spherically symmetric, there will be no explicit dependence on $m$ in the radial eigenfunctions and eigenfrequency. We can straight away remove two constants by demanding regularity at the stellar centre,
\begin{equation}
    b_l = d_l = 0\,.
\end{equation}
To determine the variables, we require six conditions. Three are provided by examining the Euler equation~\eqref{eq:Euler} in the two regions. The remaining three constraints come from the boundary conditions: continuity of $\xi^r$ and the Lagrangian perturbation of the pressure $\Delta p$ at the interface $r = r_\text{i}$, and vanishing pressure $\Delta p = 0$ at the surface $r = R$.

To make the calculation more tractable, we will assert that the density jump is small, such that
\begin{equation}
    \rho_\text{o} = \rho_\text{i} (1 - \epsilon)\,,
\end{equation}
where $|\epsilon| \ll 1$ is a small, dimensionless parameter. We will see in a moment how we require that $\epsilon$ is positive in order for there to be an oscillation due to the interface. We work to leading order in $\epsilon$ to find the solutions to the perturbation problem. The boundary condition at the surface provides
\begin{equation}
    \left( \bar{c}_l - l \bar{a}_l \frac{4 \pi G \rho_\text{i}^2}{3} \left\{ 1 + \epsilon \left[ - 2 + \left( \frac{r_\text{i}}{R} \right)^3 \right] \right\} \right) R^l + \left( \bar{d}_l + (l + 1) \bar{b}_l \frac{4 \pi G \rho_\text{i}^2}{3} \left\{ 1 + \epsilon \left[ - 2 + \left( \frac{r_\text{i}}{R} \right)^3 \right] \right\} \right) \frac{1}{R^{l + 1}} = 0\,.
    \label{eq:Condition4}
\end{equation}
We can eliminate the coefficients $c_l$, $\bar{c}_l$ and $\bar{d}_l$ using Eq.~\eqref{eq:Euler} on both sides of the interface. These terms appear in Eq.~\eqref{eq:Condition4} and the continuity of $\Delta p$ at $r = r_\text{i}$, which become
\begin{align}
    \left[ \omega^2 (1 - \epsilon) \bar{a}_l + \left( - \omega^2 + \epsilon l \frac{4 \pi G \rho_\text{i}}{3} \right) a_l \right] r_\text{i}^l + \omega^2 (1 - \epsilon) \frac{\bar{b}_l}{r_\text{i}^{l + 1}} &= 0\,, \label{eq:Condition6} \\
\begin{split}
    \left( \omega^2 - l \frac{4 \pi G \rho_\text{i}}{3} - \epsilon \left\{ \omega^2 + l \frac{4 \pi G \rho_\text{i}}{3} \left[ - 2 + \left( \frac{r_\text{i}}{R} \right)^3 \right] \right\} \right) \bar{a}_l R^l\qquad\qquad& \\
    + \left( \omega^2 + (l + 1) \frac{4 \pi G \rho_\text{i}}{3} - \epsilon \left\{ \omega^2 - (l + 1) \frac{4 \pi G \rho_\text{i}}{3} \left[ - 2 + \left( \frac{r_\text{i}}{R} \right)^3 \right] \right\} \right) \frac{\bar{b}_l}{R^{l + 1}} &= 0\,.
\end{split}\label{eq:Condition4new}
\end{align}
We are currently left with Eqs.~\eqref{eq:Condition6} and \eqref{eq:Condition4new}, as well as the continuity of $\xi^r$, which depend on $a_l$, $\bar{a}_l$ and $\bar{b}_l$. We remove $a_l$ by combining the continuity condition on $\xi^r$ and \eqref{eq:Condition6} to obtain
\begin{equation}
    \epsilon \left( \omega^2 - l \frac{4 \pi G \rho_\text{i}}{3} \right) l \bar{a}_l r_\text{i}^l = \left\{ (2 l + 1) \omega^2 - \epsilon l \left[ \omega^2 + (l + 1) \frac{4 \pi G \rho_\text{i}}{3} \right] \right\} \frac{\bar{b}_l}{r_\text{i}^{l + 1}}\,.
    \label{eq:Condition6new}
\end{equation}
Since the coefficient of $\bar{a}_l$ in Eq.~\eqref{eq:Condition6new} is $O(\epsilon)$, we can immediately infer that $\bar{b}_l = 0$ if the discontinuity vanishes. This describes the standard \textit{f}-mode behaviour, which rises as $\propto r^l$ up to the surface. In the absence of an interface, this is the only mode that a uniform-density star supports. We will encounter its solution for the star with a phase transition in a moment. Finally, we combine Eqs.~\eqref{eq:Condition4new} and \eqref{eq:Condition6new} to remove $\bar{a}_l$ or $\bar{b}_l$ from the system. Either way, we arrive at a quadratic equation in $\omega^2$,
\begin{equation}
\begin{split}
    \left( - (2 l + 1) + \epsilon \left\{ 2 l + 1 + l \left[ 1 - \left( \frac{r_\text{i}}{R} \right)^{2 l + 1} \right] \right\} \right) \omega^4& \\
    + l \left( 2 l + 1 + \epsilon \left\{ (2 l + 1) \left[ - 2 + \left( \frac{r_\text{i}}{R} \right)^3 \right] + 1 - \left( \frac{r_\text{i}}{R} \right)^{2 l + 1} \right\} \right) \frac{4 \pi G \rho_\text{i}}{3} \omega^2& \\
    - \epsilon l^2 (l + 1) \left[ 1 - \left( \frac{r_\text{i}}{R} \right)^{2 l + 1} \right] \left( \frac{4 \pi G \rho_\text{i}}{3} \right)^2& = 0\,.
\end{split}
    \label{eq:Quadratic}
\end{equation}
From the quadratic expression~\eqref{eq:Quadratic}, we will determine the oscillation frequencies for the two modes that the star supports.

The first has frequency
\begin{equation}
    \omega^2 = \frac{G M}{R^3} l\,,
\end{equation}
where $M$ represents the total mass of the star. This is the familiar result for the fundamental \textit{f}-mode of an incompressible star (in the Cowling approximation). It is interesting to note that we obtain precisely the same formula in terms of $M / R^3$ as if the interface disappeared. This provides a simple illustration of why universal relations fare so well \cite{2014PhRvD..90l4023C}. We also observe that its amplitude is (to linear order) totally insensitive to the interface, since the coefficients are
\begin{equation}
    \bar{a}_l = a_l\,, \qquad \bar{b}_l = 0\,.
\end{equation}
As we noted above, the \textit{f}-mode eigenfunctions rise gradually up the surface, which (effectively) behaves as an interface between the fluid and vacuum exterior. In this sense, the \textit{f}-mode is like an interface mode. Since the \textit{f}-mode couples so efficiently to the tide, this provides a hint for why \textit{i}-modes may also have strong tidal couplings.

The second solution to Eq.~\eqref{eq:Quadratic} is $O(\epsilon)$ and thus arises due to the presence of the interface in the fluid interior. This mode oscillates at
\begin{equation}
    \omega^2 = \epsilon \frac{G M}{R^3} \frac{l (l + 1)}{2 l + 1} \left[ 1 - \left( \frac{r_\text{i}}{R} \right)^{2 l + 1} \right]\,.
    \label{eq:omega2}
\end{equation}
This is the interfacial \textit{i}-mode of the star that is sourced by the presence of the density discontinuity $\epsilon$. Here, we see that $\epsilon$ must be positive to give rise to a real, oscillating solution. Otherwise, the perturbation is immediately damped. The eigenfunctions of the \textit{i}-mode are given by
\begin{equation}
    \bar{a}_l = - a_l \frac{1}{(R / r_\text{i})^{2 l + 1} - 1} (1 + \epsilon)\,, \qquad \bar{b}_l = - a_l \frac{l}{l + 1} \frac{R^{2 l + 1}}{(R / r_\text{i})^{2 l + 1} - 1} \left[ 1 + \epsilon \left( \frac{r_\text{i}}{R} \right)^{2 l + 1} \right]\,.
\end{equation}
Here, we see that this mode rises in an identical fashion to the \textit{f}-mode up to the interface and then obtains a non-vanishing $\bar{b}_l$ that decays as $\propto 1 / r^{l + 1}$. It is interesting to note that in the $\epsilon \rightarrow 0$ limit, the \textit{i}-mode becomes a trivial current in the star: both $\bar{a}_l$ and $\bar{b}_l$ are non-zero in this limit. Only when $\epsilon > 0$, the mode starts oscillating and sourcing finite density and pressure perturbations.

We will now examine the tidal overlap integrals of these two modes. The overlap is determined by \cite{1994MNRAS.270..611L}
\begin{equation}
\begin{split}
    Q_l &= \int \delta \rho_l r^{l + 2} \, dr \\
    &= \rho_\text{i} \left\{ l \bar{a}_l \left[ R^{2 l + 1} - \epsilon \left( R^{2 l + 1} - r_\text{i}^{2 l + 1} \right) \right] - (l + 1) \bar{b}_l \right\}\,,
\end{split}
\end{equation}
where $\delta \rho(r, \theta, \varphi) = \delta \rho_l(r) Y_l^m(\theta, \varphi)$. We observe that $Q_l$ is the contribution of the mode to the mass multipole moment of degree $l$. Next, we note that the oscillating solutions we have found have free amplitudes. However, when the tide sources the perturbation, this amplitude is set by the tidal coupling. Therefore, it is necessary to introduce the mode normalisation constant
\begin{equation}
\begin{split}
    \mathcal{A}^2 &= \int \rho \left[ \left( \frac{d\psi_l}{dr} \right)^2 r^2 + l (l + 1) \psi_l^2 \right] \, dr \\
    &= \rho_\text{i} \left\{ l a_l^2 r_\text{i}^{2 l + 1} + (1 - \epsilon) \left[ 1 - \left( \frac{r_\text{i}}{R} \right)^{2 l + 1} \right] \left[ l \bar{a}_l^2 R^{2 l + 1} + (l + 1) \frac{\bar{b}_l^2}{r_\text{i}^{2 l + 1}} \right] \right\}\,.
\end{split}
\end{equation}
The choice of normalisation does not impact the result. It is the quantity $Q_l / \mathcal{A}$, which is independent of the amplitude, that describes the dynamics \cite{2020PhRvD.101h3001A}.

Now, we calculate the couplings for the two mode solutions. The overlap between the tide and the \textit{f}-mode is particularly simple and we find that
\begin{equation}
    \frac{Q_l}{\mathcal{A} / \sqrt{M R^2}} = M R^l \sqrt{\frac{3 l}{4 \pi}} \left\{ 1 - \epsilon \frac{1}{2} \left[ \left( \frac{r_\text{i}}{R} \right)^3 - \left( \frac{r_\text{i}}{R} \right)^{2 l + 1} \right] \right\}\,.
\end{equation}
In contrast to its eigenfrequency, the overlap integral for the \textit{f}-mode does have a dependence on $\epsilon$ and $r_\text{i}$. The calculation for the \textit{i}-mode is substantially more involved, requiring that the mode eigenfunctions are determined to second order in $\epsilon$. Therefore, not an insignificant amount of algebra reveals
\begin{equation}
    \frac{Q_l}{\mathcal{A} / \sqrt{M R^2}} = - \epsilon^2 M R^l \sqrt{\frac{3}{4 \pi} \frac{l (l + 1)}{2 l + 1}} \sqrt{1 - \left( \frac{r_\text{i}}{R} \right)^{2 l + 1}} \left( \frac{r_\text{i}}{R} \right)^{(2 l + 1) / 2} \left[ 1 - \left( \frac{r_\text{i}}{R} \right)^3 \right]\,.
    \label{eq:Ql}
\end{equation}
Here, we see that the overlap for the \textit{i}-mode is also dependent on the depth of the phase transition and its position in the star. The overlap integral is a function of $\epsilon^2$ and has a non-linear dependence on the position of the interface. For $l = 2$, we find that the overlap has a maximum at $r_\text{i} \approx 0.726 R$.

\section{Mode calculation and detectability results}

In order to streamline the presentation in the main text, we have focused on the key message: The anticipated detectability of the interface modes associated with a first-order phase transition. Here, we provide some additional information that may be useful to an interested reader.

Let us first discuss the perturbation calculation. Our mode calculation is functionally identical to that of Ref.~\cite{2025MNRAS.536.1967C}. To begin with, we must establish the background model, which is taken to have the usual Schwarzschild form
\begin{equation}
    ds^2 = - e^\nu \, c^2 dt^2 + e^\lambda \, dr^2 +r^2 (d\theta^2 + \sin^2 \theta \, d\varphi^2)
\end{equation}
and be described by the perfect-fluid stress-energy tensor
\begin{equation}
    T^{a b} = \frac{1}{c^2} (\varepsilon + p) u^a u^b + p \, g^{a b} \,,
\end{equation}
where $u^a$ is the fluid four velocity and $g^{a b}$ the inverse of the metric $g_{a b}$.
These assumptions about the spacetime and matter mean that the background configuration is obtained by solving the standard Tolman-Oppenheimer-Volkoff equations.

To determine the oscillation modes, we use linear perturbation theory.
The linear perturbations of a relativistic star can be described by the Eulerian perturbation of the metric $h_{a b}$ and the Lagrangian displacement vector of the fluid $\xi^a$, we then assume a gauge condition and only consider its spatial components $\xi^i$.
We choose to simplify the problem by introducing the (relativistic) Cowling approximation. 
We take this to mean that we ignore the perturbations of the gravitational field $h_{a b} = 0$. 

As we are considering non-rotating stars, we assume that the oscillation modes, with frequency $\omega$, are associated with a polar displacement vector, which in spherical polar coordinates $(r, \theta, \varphi)$ is given by
\begin{equation}
\begin{split}
    \xi^r(r, \theta, \varphi) &= \frac{1}{r} W_l(r) \, Y_l^m(\theta, \varphi)\,, \\
    \xi^\theta(r, \theta, \varphi) &= \frac{1}{r^2} V_l(r) \, \partial_\theta Y_l^m(\theta, \varphi)\,, \\
    \xi^\varphi(r, \theta, \varphi) &= \frac{1} {r^2 \sin^2 \theta} V_l(r) \, \partial_\varphi Y_l^m(\theta, \varphi)\,,
\end{split}
\end{equation}
where $W_l$ and $V_l$ are the radial and angular amplitudes, respectively, and $Y_l^m$ is a spherical harmonic with multipolar order $l$ and azimuthal degree $m$.
To find an oscillation mode, one must solve a pair of coupled differential equations for $W_l$ and $V_l$, with appropriate boundary conditions. The system of equations we solve are those provided in Ref.~\cite{2025MNRAS.536.1967C}.

What is different in this calculation is the presence of phase transitions in the assumed equations of state. Our strategy---unlike previous approaches, such as Ref.~\cite{2024ApJ...964...31M}---is to integrate through the discontinuities, allowing the numerics to resolve the transition. We find that our method is sufficient to locate the interface modes due to these phase transitions.

Now, we comment on the identification of the interface modes. As is already clear from the incompressible model calculated above, the eigenfunctions of an interface mode tend to exhibit a sharp peak at the local of the interface. An illustration of this is provided in Fig.~1 in the main text, which shows the eigenfunctions $W_l(r)$ and $V_l(r)$ for the \textit{i}-mode of a $1.4 M_\odot$ star using the equation of state labelled 461 in Fig.~1. The dominant features are a discontinuity in $V_l$ and a discontinuity in the derivative of $W_l$, both located at the phase transition. This behaviour is noticeably distinct from the other common modes in the oscillation spectrum (\textit{f}-, \textit{p}-, \textit{g}-modes \textit{etc.}). 

\begin{figure}
    \includegraphics[width=0.5\textwidth]{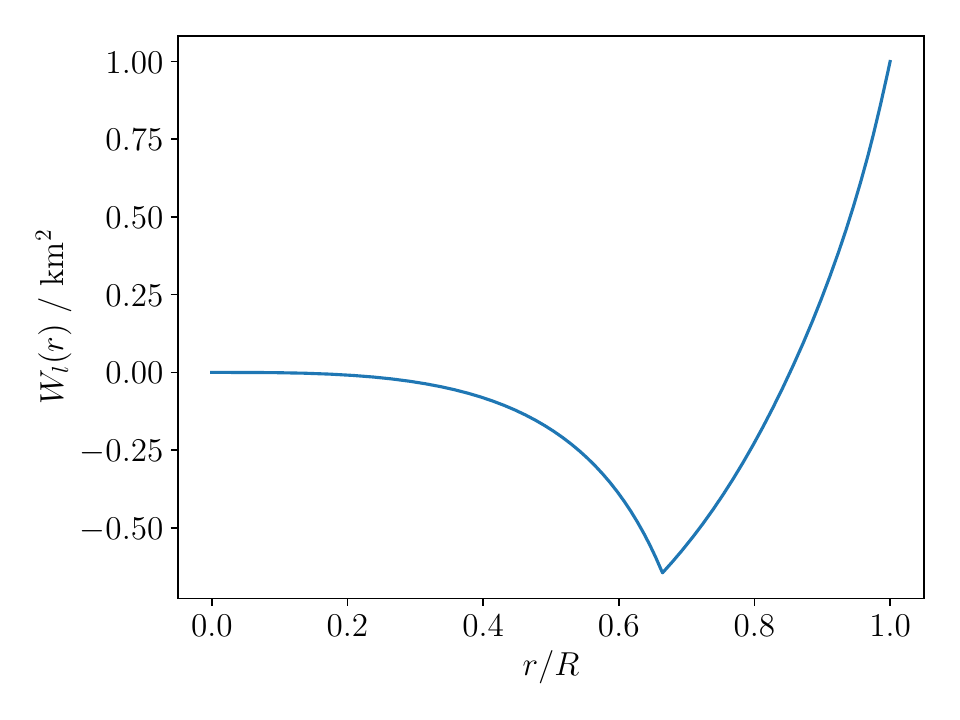}%
    \includegraphics[width=0.5\textwidth]{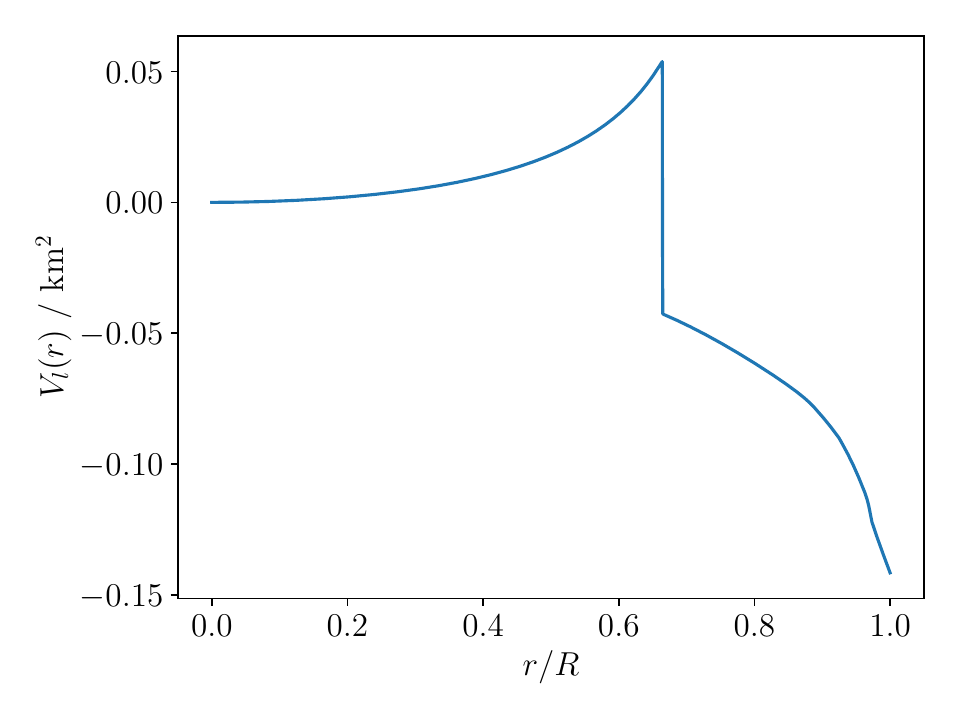}
    \caption{\label{fig:IModeVW}%
    Eigenfunctions $W_l(r)$ and $V_l(r)$ for a typical interface mode of an $M = 1.4 M_\odot$ neutron star. As explored in the incompressible stellar model, the eigenfunctions rise up to the location of the interface and then inherit a sharp kink. The radial eigenfunction $W_l$ is continuous across the interface, while the tangential eigenfunction $V_l$ is not.}
\end{figure}

\begin{figure}
    \centering
    \includegraphics[width=0.5\textwidth]{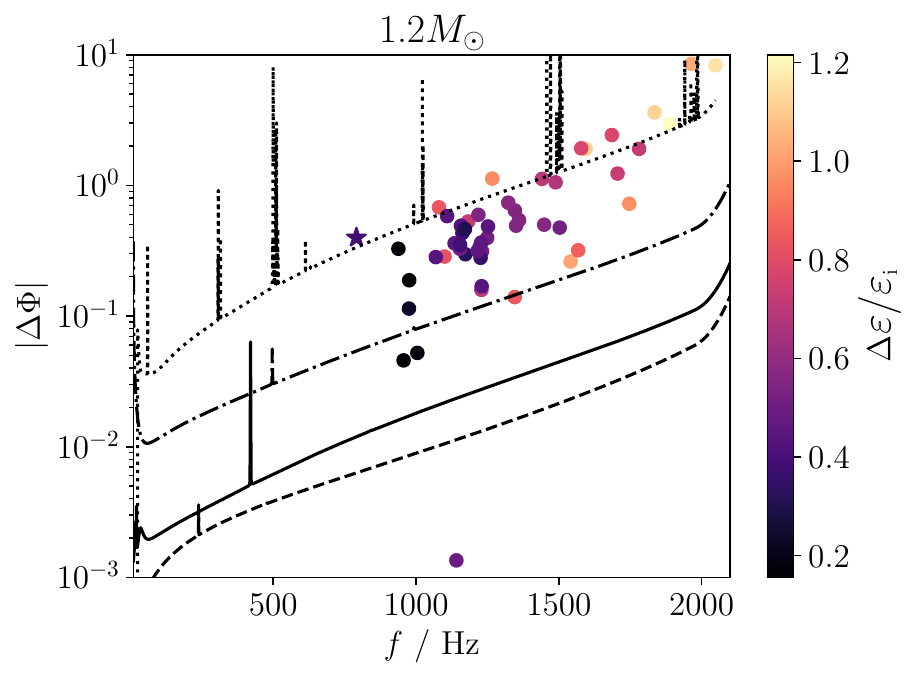}%
    \includegraphics[width=0.5\textwidth]{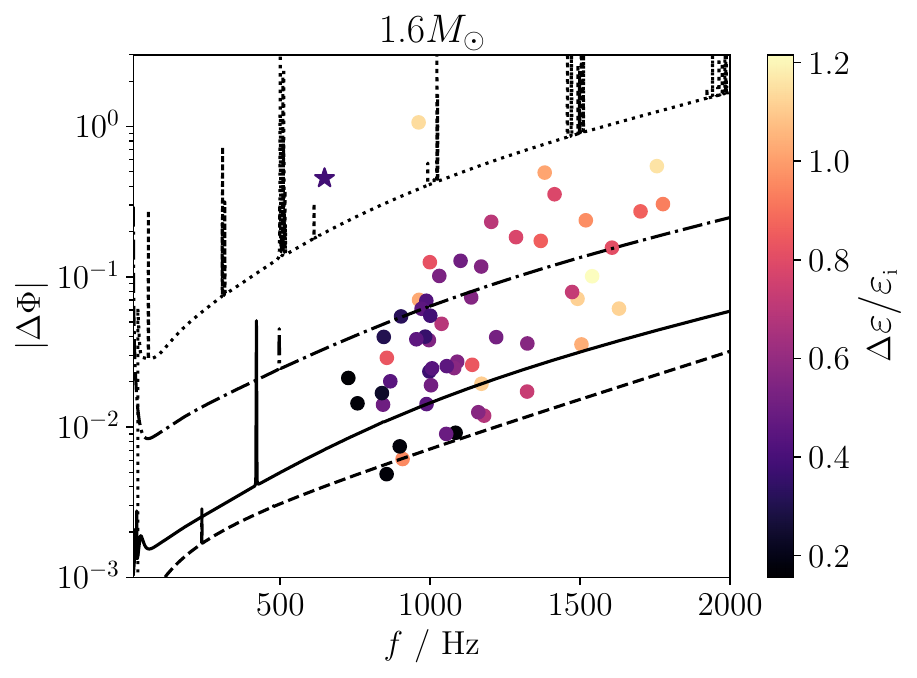}
    \includegraphics[width=0.5\textwidth]{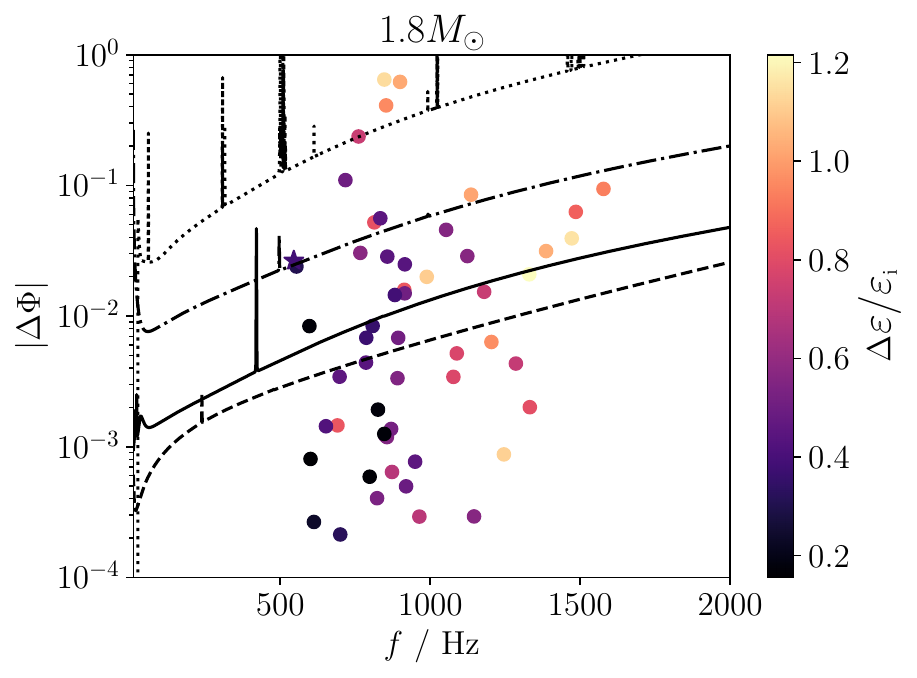}
    \caption{\label{fig:OtherDetectabilityTrendsEps}%
    Estimated shift in orbital phase $|\Delta \Phi|$ against gravitational-wave frequency $f$ for equal-mass $M = 1.2 M_\odot$, $1.6 M_\odot$ and $1.8 M_\odot$ binaries.
    The markers, curves and colour bar are described in Fig.~2, which this figure complements by illustrating the scaling with mass.}
\end{figure}

\begin{figure}
    \includegraphics[width=0.5\textwidth]{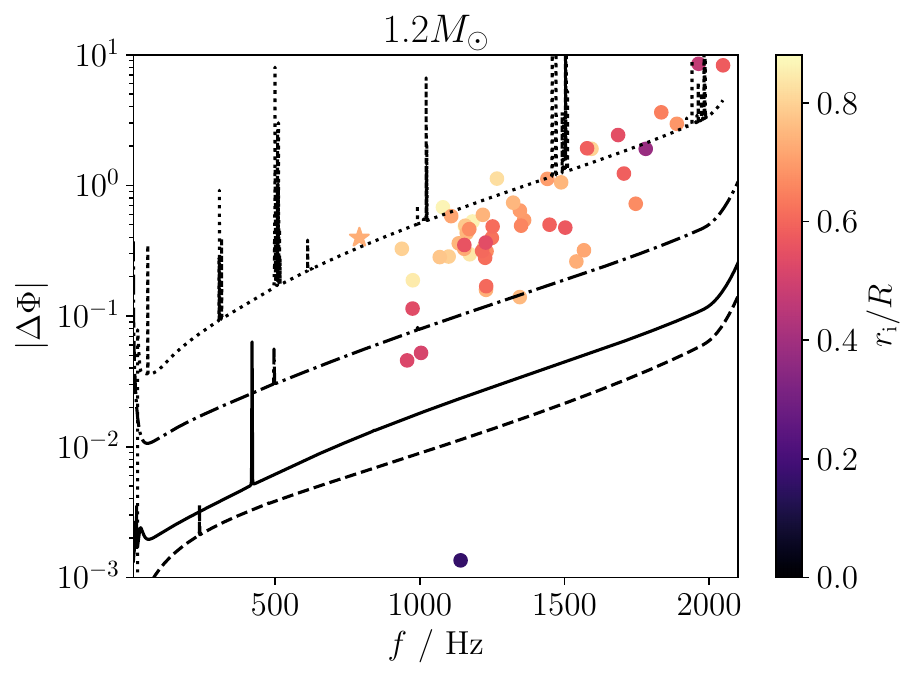}%
    \includegraphics[width=0.5\textwidth]{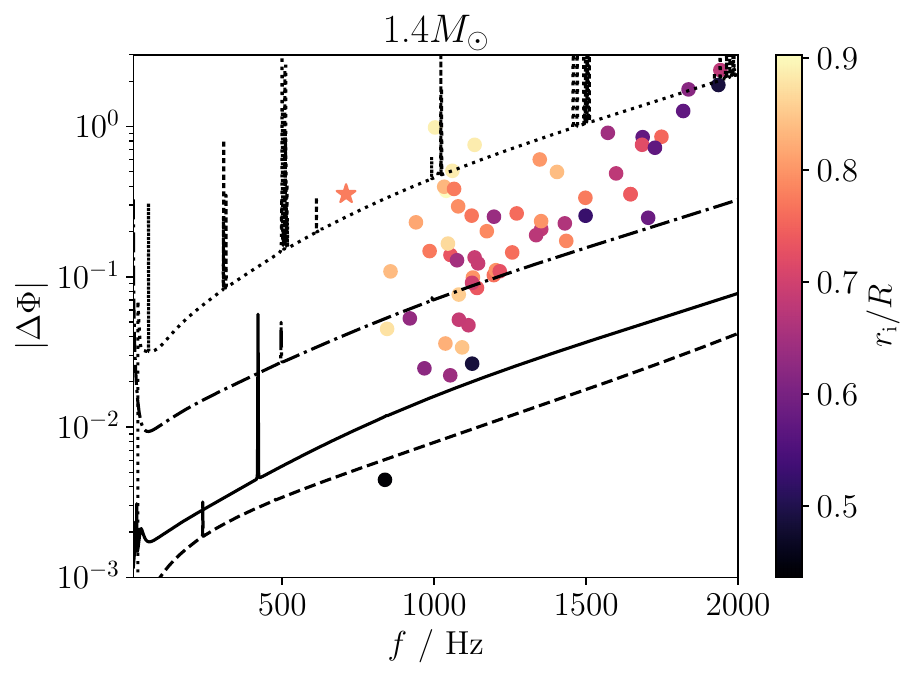}
    \includegraphics[width=0.5\textwidth]{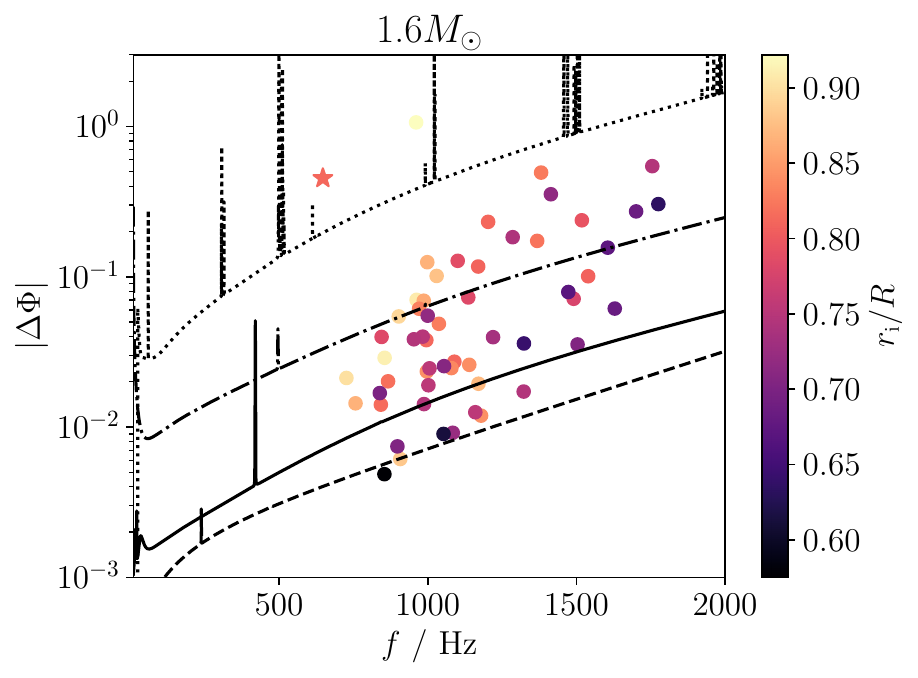}%
    \includegraphics[width=0.5\textwidth]{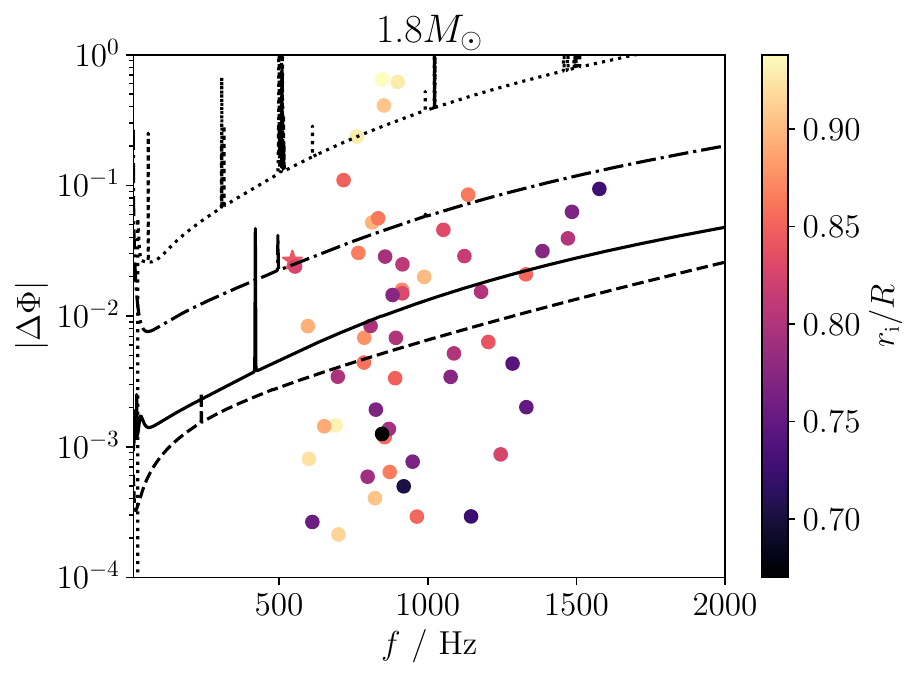}
    \caption{\label{fig:DetectabilityTrendsR0}%
    Estimated shift in orbital phase $|\Delta \Phi|$ against gravitational-wave frequency $f$ for equal-mass $M = 1.2 M_\odot$, $1.4 M_\odot$, $1.6 M_\odot$ and $1.8 M_\odot$ binaries.
    The markers and curves are described in Fig.~2, which this figure complements by illustrating the scaling with mass.
    The colour bar shows the location of the phase transition $r_i / R$ for each star.
    As one might expect from the incompressible-model problem in Sec.~\ref{app:Model}, it is not straightforward to identify an overall trend from these results.}
\end{figure}

Next, there are two natural questions one may ask about the results presented in Fig.~2 in the main text.  First, noting that the tidal interaction scales with the masses involved, one may want to consider how the results change for different neutron-star masses. As an illustration of this, we provide results also for (equal mass) $M = 1.2 M_\odot$, $1.6 M_\odot$ and $1.8 M_\odot$ systems in Fig.~\ref{fig:OtherDetectabilityTrendsEps}. The second questions concerns how the results depend on the location of the phase transition in the star. It should be evident from the above analytical calculation that this behaviour is somewhat less obvious. Nevertheless, we provide the relevant results in Fig.~\ref{fig:DetectabilityTrendsR0}.

